\documentclass[showpacs,twocolumn,aps,pra]{revtex4}
\usepackage{mathrsfs}
\usepackage{amsmath}
\usepackage{amsfonts}
\usepackage{amssymb}
\usepackage{amsthm}
\usepackage{graphicx}
\usepackage{color}
\usepackage{epsfig}
\usepackage{cancel}
\usepackage{bm}
\usepackage{sidecap}
\usepackage{epstopdf}

\providecommand{\U}[1]{\protect\rule{.1in}{.1in}}
\usepackage{ulem}
\begin{document}
\title{Chiral anomaly of Weyl magnons in stacked honeycomb ferromagnets}
\author{Ying Su$^{1,2}$}
\author{X. R. Wang$^{1,2}$}
\email[corresponding author: ]{phxwan@ust.hk}
\affiliation{$^{1}$Physics Department, The Hong Kong University of
Science and Technology, Clear Water Bay, Kowloon, Hong Kong}
\affiliation{$^{2}$HKUST Shenzhen Research Institute, Shenzhen 518057,
China}
\date{\today}
\begin{abstract}
Chiral anomaly of Weyl magnons (WMs), featured by nontrivial band crossings 
at paired Weyl nodes (WNs) of opposite chirality, is investigated. 
It is shown that WMs can be realized in stacked honeycomb ferromagnets. 
Using the Aharonov-Casher effect that is about the interaction between 
magnetic moments and electric fields, the magnon motion in honeycomb 
layers can be quantized into magnonic Landau  levels (MLLs). 
The zeroth MLL is chiral so that unidirectional WMs propagate in the 
perpendicular (to the layer) direction for a given WN under a magnetic 
field gradient from one WN to the other and change their chiralities, 
resulting in the magnonic chiral anomaly (MCA). 
A net magnon current carrying spin and heat through the zeroth MLL 
depends linearly on the magnetic field gradient and 
the electric field gradient in the ballistic transport. 
\end{abstract}

\vspace{-0.2in}
\maketitle
Topological magnetic states have attracted enormous attention in 
recent years \cite{TM1,TM2,TM3,TM4,TM5,TM6,TM7,TM8,TM9,TM10,TM11,TM12,
TM13,TM14,TM15,TM16,TM17,DM,WM1,WM2,WM3,NL,TM19}, because of their 
fundamental interest and importance in magnonics that is about 
generation, detection, and manipulation of magnons \cite{M1,M2,M3,M4}.
Magnons, the quanta of low-energy excitations of magnetic materials, can 
carry, process, and transmit information \cite{TS1,TS2} like electrons besides 
being a control knob of magnetization dynamics \cite{yanpeng,hubin,xiansi}.
So far, almost all studies on Weyl magnons (WMs) focus on the nontrivial 
band topology of Weyl nodes (WNs) and magnon surface states in pyrochlore
magnetic materials \cite{WM1,WM2,WM3}. There is no study of  
magnonic chiral anomaly (MCA), one of the two important signatures of Weyl 
materials [the other is the magnon (Fermi) arc on sample surfaces due 
to topologically protected surface states between two paired WNs]. 
The realization and detection of the MCA are the main theme of this work. 

To study the MCA, one needs to have three-dimensional (3D) Weyl magnetic 
materials and to realize the magnonic Landau levels (MLLs) first. 
Only then one can study the magnon transport through the zeroth MLL under 
driving forces, i.e. the MCA. WNs in Weyl magnetic materials appear 
usually at high energy \cite{WM1,WM2,WM3}. Thus, it is not trivial to 
inject magnons into the high-energy WNs to creat WMs. 
In this paper, we show that stacked honeycomb ferromagnets can be a 
Weyl magnetic material that supports both type-I WMs and type-II WMs. 
A two-band model is found with either one pair or two pairs of WNs. 
Magnons can interact with electric fields through the Aharonov-Casher 
effect \cite{AC} so that magnon motion in honeycomb layers can be 
quantized into the MLLs by a proper inhomogeneous electric field. 
A quasi-one-dimensional magnon conductor connected to two magnon 
reservoirs under a proper inhomogeneous magnetic field perpendicular 
to the honeycomb layers is used to study the MCA and magnon transport 
in the longitudinal direction. 

\begin{figure}
  \begin{center}
  \includegraphics[width=8.5 cm]{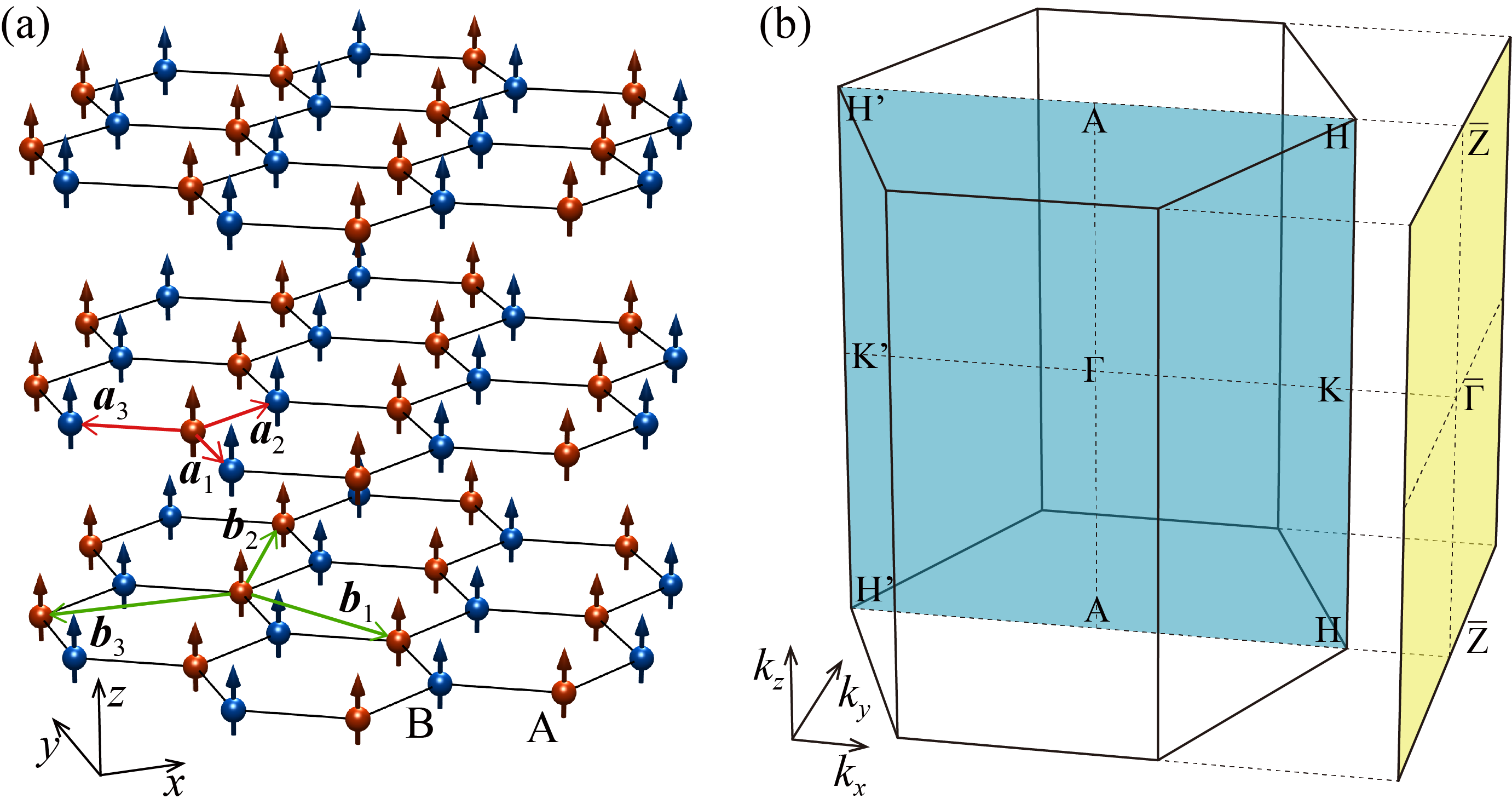}
  \end{center}
  \vspace{-0.2in}
\caption{(color online). \footnotesize{(a) Schematic diagram of 
stacked honeycomb ferromagnets. Each site has a spin of $\bm S$ polarized in 
the $z$ direction, represented by a red dot with an arrow on sublattice A 
and a blue dot with an arrow on sublattice B. The in-plane vectors 
$\bm{a}_{1(2,3)}$ and $\bm{b}_{1(2,3)}$ are explained in the text. 
(b) The first bulk Brillouin zone (BZ) and the first (100) surface 
BZ of the stacked honeycomb ferromagnets.}}
  \label{fig1}
  \vspace{-0.2in}
\end{figure}

Our 3D WMs exist in stacked honeycomb ferromagnets as shown in 
Fig.~\ref{fig1}(a). The honeycomb lattices (in the $xy$ plane) 
are perfectly aligned in the $z$ direction. A spin $\bm{S}$ 
(in units of $\hbar$) polarized in the $z$ direction is on each site. 
$\bm{a}_i$ (red arrows) and $\bm{b}_i$ (green arrows) ($i=1,2,3$) defined 
in honeycomb layers are three vectors connecting nearest neighbor (NN) 
sites and three vectors connecting next nearest neighbor (NNN) sites, 
respectively \cite{ab}. A and B denote two sublattices of honeycomb layers. 
The layer separation is the same as the distance between two NN 
intralayer lattice sites that is set as unity. 
The spin Hamiltonian reads 
\begin{equation}
\begin{split}
&H=-J\sum_{\langle i,j\rangle,l}\bm{S}_{i,l}\cdot\bm{S}_{j,l} - 
\sum_{i,l} K_iS_{i,l}^{z2}  - \sum_{i,\langle l,l'\rangle} 
J_i \bm{S}_{i,l}\cdot \bm{S}_{i,l'}\\
&+ D\sum_{\langle\langle i,j\rangle\rangle,l} \nu_{ij} \hat{\bm{z}}
\cdot(\bm{S}_{i,l}\times\bm{S}_{j,l})- g\mu_B B_0\sum_{i,l} S_{i,l}^z , 
\end{split}
\label{H1}
\end{equation}
where $i$ and $j$ label lattice sites in honeycomb layers, and $l$ and $l'$ 
are layer indexes. $\langle i,j\rangle$ and $\langle\langle i,j\rangle\rangle$ 
denote the NN and NNN intralayer sites, and $\langle l,l'\rangle$ are the NN 
layers. The first term describes the NN intralayer ferromagnetic exchange 
interaction with $J>0$. The second term is the anisotropy energy with easy 
axis along $z$ direction and anisotropy coefficients $K_i=K_{\text{A}}$ 
($K_{\text{B}}$) for sites on sublattice A (B). 
$J_i=J_{\text{A}}$ ($J_{\text{B}}$) in the third term are the NN interlayer 
exchange coefficients for sites on sublattice A (B). The fourth term 
describes the Dzyaloshinskii-Moriya interaction \cite{TM14,TM15,TM16,TM17,DMI} 
and $\nu_{ij}=(2/\sqrt{3})(\hat{\bm{d}}_1\times\hat{\bm{d}}_2)_z=\pm 1$, where 
$\hat{\bm{d}}_1$ and $\hat{\bm{d}}_2$ are the unit vectors along NN intralayer 
bonds connecting the common NN site of $i$ and $j$ to site $j$ and $i$. 
The last term is the Zeeman energy due to the external magnetic field $B_0$ 
along $z$ direction ($B_0=0$ is assumed below since it only shifts the energy). 

Under the Holstein-Primakoff transformation 
$S^+_{i,l}= \sqrt{2S-n_{i,l}} c_{i,l}$, $S^-_{i,l}= c_{i,l}^\dagger 
\sqrt{2S-n_{i,l}}$, $n_{i,l}=c_{i,l}^\dagger c_{i,l}$ \cite{HPT},
where $c_{i,l}^\dagger$ and $c_{i,l}$ are magnon creation and annihilation
operators satisfying the boson commutation relations, 
Hamiltonian (\ref{H1}) becomes a tight-binding Hamiltonian 
\begin{equation}
\begin{split}
&H=-JS\sum_{\langle i,j\rangle,l}(c_{i,l}^\dagger c_{j,l}+\text{H.c.})-
DS\sum_{\langle\langle i,j\rangle\rangle,l}(i\nu_{ij}c_{i,l}^\dagger 
c_{j,l} \\ 
&+ \text{H.c.} )+ \sum_{i,l} V_i c_{i,l}^\dagger c_{i,l} - 
\sum_{i,\langle l,l'\rangle} J_i S (c_{i,l}^\dagger c_{i,l'}+\text{H.c.}), 
\end{split}
\label{H2}
\end{equation}
where $V_i=3JS+2K_i S+2J_i S$ is the sublattice-dependent on-site energy. 
The Hamiltonian can be block diagonalized in momentum space as  
$H=\sum_{\bm{k}}c_{\bm{k}}^\dagger \mathcal{H}(\bm{k})c_{\bm{k}},$ where 
$c_{\bm k}=(a_{\bm k},b_{\bm k})^{\rm T}$ and $\mathcal{H}({\bm k})=
\varepsilon_0(\bm{k})I+\sum_\beta h_\beta(\bm{k})\sigma_\beta$, ($\beta=x,y,z$). 
$a_{\bm k}$ and $b_{\bm k}$ are respectively defined on sublattices A and B. 
$I$, $\sigma_x$, $\sigma_y$, and $\sigma_z$ are the $2\times 2$ 
identity matrix, and three Pauli matrices. The other quantities are 
$\varepsilon_0(\bm{k})=3JS+K_+S+J_+S(1-\cos k_z)$,
$h_x(\bm{k})=-JS\sum_{j=1}^3 \cos(\bm{k}\cdot\bm{a}_j)$,
$h_y=-JS\sum_{j=1}^3 \sin(\bm{k}\cdot\bm{a}_j)$, 
$h_z(\bm{k})=2DS\sum_{j=1}^3\sin(\bm{k}\cdot\bm{b_j})+K_-S+J_-S(1-\cos k_z)$,  
$K_\pm=K_\text{A}\pm K_\text{B}$ and $J_\pm=J_\text{A}\pm J_\text{B}$. 
The dispersion relations of the two magnon bands are $\varepsilon_\pm
(\bm{k})=\varepsilon_0(\bm{k})\pm\sqrt{\sum_\beta h_\beta^2(\bm{k})}$. 

\begin{figure}
  \begin{center}
  \includegraphics[width=8.5 cm]{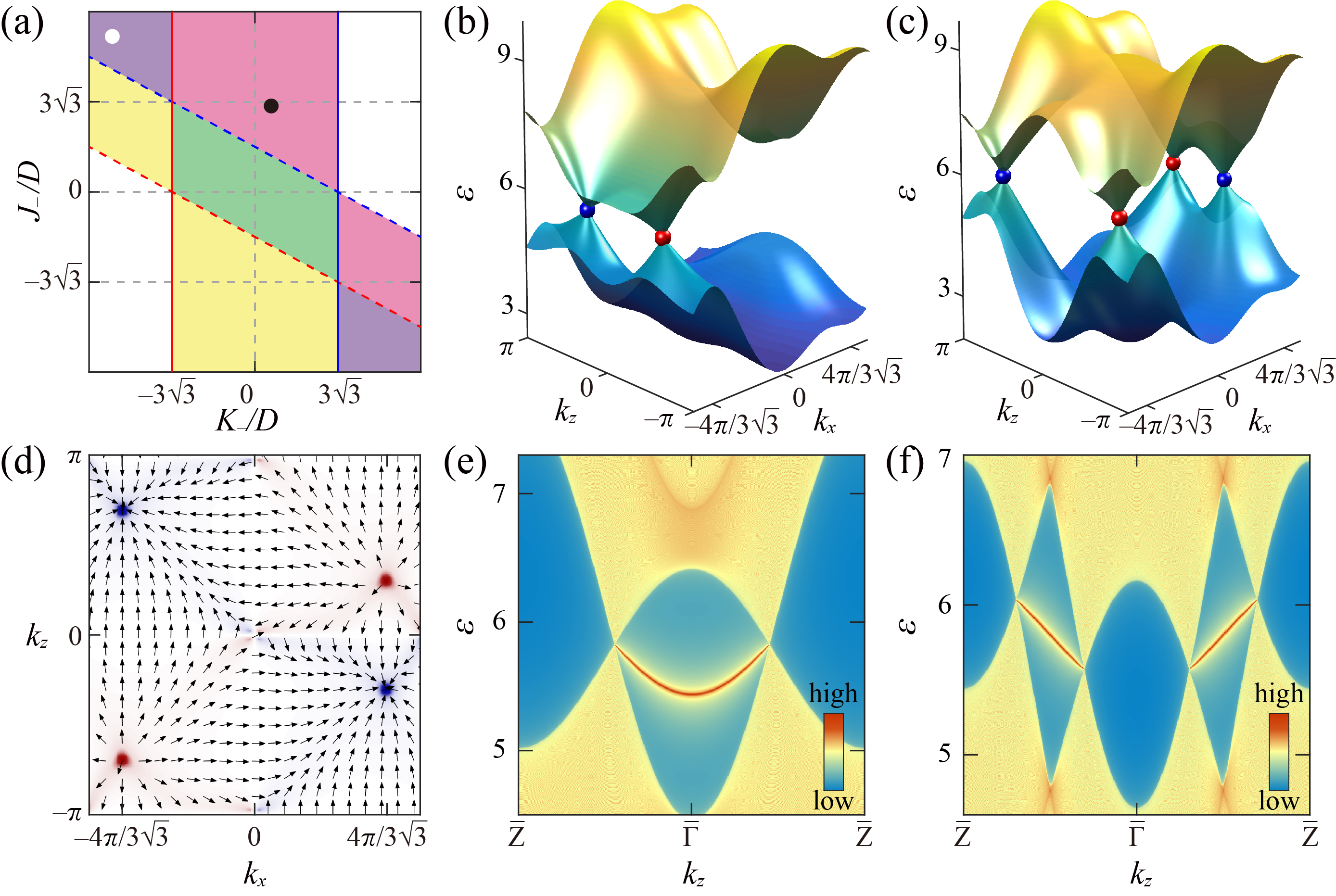}
  \end{center}
  \vspace{-0.2in}
\caption{(color online). \footnotesize{(a) Phase diagram in the 
$K_-/D$-$J_-/D$ plane divided by four boundaries $K_-=\pm 3\sqrt{3}D$ 
and $K_-+2J_-=\pm3\sqrt{3}D$. The yellow, pink, and purple regions 
denote three different WM phases. The system is in the topological 
magnon insulator phase in the green region while it is the trivial 
gapped phase in the white region. The black and white dots correspond 
to the magnon band structures in (b) and (c) whose model parameters 
are specified in the text. 
The Weyl nodes of chirality $\pm 1$ are marked by red and blue dots. 
(d) The Berry curvature of the lower magnon band in (c). 
The arrows represent the direction of Berry curvature vectors in 
the $k_xk_z$ plane with $k_y=0$. The background color denotes the 
divergence of Berry curvature, where red and blue represent positive 
and negative values, respectively. (f) and (e) The density plot of 
(100) surface spectral functions for the magnon bands in (b) and (c). 
Enegy $\varepsilon$ is in units of $JS$, and insets are the colorbars.}} 
  \label{fig2}
  \vspace{-0.2in}
\end{figure}

The two magnon bands cross at $\bm{k}$ called a WN when $h_{x(y,z)}(\bm{k})=0$. 
This may happen on two lines, HKH and H'K'H', in the first Brillouin 
zone (BZ) shown in Fig.~\ref{fig1}(b) because $h_{x(y)}(\bm{k})=0$ 
and $h_z(\bm{k})=\pm3\sqrt{3}DS+K_-S+J_-S(1-\cos k_z)$ for 
$\bm{k}=(\pm 4\pi/3\sqrt{3},0,k_z)$. WNs are located at 
$\bm{k}_1^\pm=(-4\pi/3\sqrt{3},0,\pm\cos^{-1}f_1)$ on H'K'H' and at 
$\bm{k}_2^\pm=(4\pi/3\sqrt{3},0,\pm\cos^{-1}f_2)$ on HKH, where 
$f_\eta = K_-/J_-+1+(-1)^\eta 3\sqrt{3}D/J_-$ ($\eta=1,2$). 
Conditions of $|f_\eta|\le 1$ result in four phase boundary lines of 
$K_-=\pm 3\sqrt{3}D$ (solid lines) and $K_-+2J_-=\pm3\sqrt{3}D$ 
(dashed lines) in Fig.~\ref{fig2}(a), that divide the 
$J_-/D$-$K_-/D$ plane into nine regions (colored differently). 
The two magnon bands are gapped in both white and green regions. 
However, the system in the green region is in a topologically 
nontrivial phase in which topologically protected surface states 
exist in the band gap (see Supplemental Material Ref. \cite{sm}). 
This phase is called topological magnon insulator in the literature \cite{TM5}. 
For an arbitrary $k_z$, Hamiltonian 
$\mathcal{H}_{k_z}(k_x,k_y)$ gives a nonzero Chern number \cite{sm}. 
On the other hand, the system in the white regions 
is in a trivial phase with zero topological number. 

The rest of regions in Fig.~\ref{fig2}(a) belongs to three different WM 
phases: the WM in the pink regions has one pair of WNs at $\bm{k}_1^\pm$; 
WMs in the yellow regions has one pair of WNs at $\bm{k}_2^\pm$; 
and WMs in the purple regions has two pairs of 
WNs at $\bm{k}_1^\pm$ and $\bm{k}_2^\pm$. 
The effective Weyl Hamiltonian (to the first order in the momentum 
deviation $\bm{q}=\bm{k}-\bm{k}_{\eta}^{\pm}$) around the WNs 
$\bm{k}_\eta^\pm$ can be obtained from the Taylor expansion as
\begin{equation}
\mathcal{H}_{\eta}^\pm(\bm{q})=(\varepsilon_{\eta}+ \hbar u_{\eta,z}^\pm 
q_z)I + \sum_{\beta=x,y,z}\hbar v_{\eta,\beta}^\pm q_{\beta}\sigma_\beta , 
\label{H3}
\end{equation}
where $\varepsilon_\eta = 3JS+K_+S+J_+S(1-f_\eta)$, 
$u_{\eta,z}^\pm=\pm J_+S\sqrt{1-f_\eta^2}/\hbar$, 
$v_{\eta,x}^\pm=(-1)^\eta 3JS/2\hbar$, $v_{\eta,y}^\pm=3JS/2\hbar$, 
and $v_{\eta,z}^\pm=\pm J_-S\sqrt{1-f_\eta^2}/\hbar$. 
The chirality of WNs at $\bm{k}_\eta^\pm$ can be calculated from the 
effective Weyl Hamiltonian Eq.~(\ref{H3}) as $\chi_\eta^\pm={\rm sgn}
(\prod_\beta v_{\eta,\beta}^\pm)=\pm(-1)^\eta{\rm sgn}(J_-)$ \cite{C}. 
Thus WNs appear in pairs with opposite chirality as required by the 
no-go theorem \cite{NN,WAN,C}. According to the classification of Weyl 
semimetals \cite{T2WM}, the WMs are of type-II when $|u_{\eta,z}^\pm|>
|v_{\eta,z}^\pm|\Rightarrow|J_+|>|J_-|$ (see Supplemental Material Ref.~\cite{sm}).
Namely, $J_{\rm A}$ and $J_{\rm B}$ have the same sign. 
Otherwise the WMs are of type-I. 
Two WNs $\bm{k}_1^\pm$ merge at K' on boundary $K_-=3\sqrt{3}D$ (blue 
solid line) and at H' on boundary $K_-+2J_-=3\sqrt{3}D$ (blue dash line), 
while two WNs $\bm{k}_2^\pm$ merge at K on boundary $K_-=-3\sqrt{3}D$ (red solid 
line) and at H on boundary $K_-+2J_-=-3\sqrt{3}D$ (red dash line).  
At $(K_-/D, J_-/D)=(-3\sqrt{3},0)$ and $(3\sqrt{3},0)$, WMs become nodal-line 
magnons \cite{NL} in which two magnon bands cross on HKH line for the former 
and on H'K'H' line for the later (see Supplemental Material Ref.~\cite{sm}).

One can use energy surfaces of two magnon bands, say as a function of 
$k_x$ and $k_z$ for fixed $k_y=0$ [the blue plane in Fig.~\ref{fig1}(b)], 
to visualize the WNs of WMs identified above. To be concrete and without 
losing generality, we set $D=0.2J$, $K_+=12D$, and $J_+=2D$ below. 
The magnon bands for $(K_-,J_-)=(D,5D)$ and $(-9D,9D)$ [respectively 
marked by the black and white dots in Fig.~\ref{fig2}(a)] are plotted 
in Fig.~\ref{fig2}(b) and 2(c). Apparently, they are type-I WMs with 
one or two pairs of WNs denoted by the red and blue dots for chirality 
$\pm1$, respectively. The magnon bands for type-II and nodal-line 
magnons are shown in the Supplemental Material Ref.~\cite{sm}. 
The WM phase can also be confirmed by the topological number 
calculations from the Berry curvature $\bm{\Omega}_\pm(\bm{k})=i\nabla_
{\bm{k}}\times\langle\pm,{\bm{k}}|\nabla_{\bm{k}}|\pm,{\bm{k}}\rangle$, 
where $|\pm,{\bm{k}}\rangle$ are the eigenstates of the upper and 
lower magnon bands, and $\bm{\Omega}_+(\bm{k})=-\bm{\Omega}_-(\bm{k})$. 
From the effective Weyl Hamiltonian Eq.~(\ref{H3}), the Berry curvature 
of the lower magnon band around the WN at $\bm{k}_\eta^\pm$ can be 
analytically calculated 
\begin{equation}
\bm{\Omega}_-(\bm{k}_\eta^\pm+\bm{q})=\frac{\prod_\beta v_{\eta,\beta}^\pm \bm{q}}
{2\left(\sum_\beta v_{\eta,\beta}^{\pm2} q_\beta^2\right)^{3/2}},
\end{equation}
which diverges at the WN (where $\bm{q}=0$), corresponding to a magnetic 
monopole there in the momentum space as shown in Fig.~\ref{fig2}(d) that 
is the numerical result of the exact Hamiltonian $\mathcal{H}(\bm{k})$ 
for the lower magnon band in Fig.~\ref{fig2}(c) with $k_y=0$.
The black arrows show the directions of the projection of 
Berry curvature onto the $k_xk_z$ plane and the color 
represents the divergence of Berry curvature $\nabla_{\bm{k}}
\cdot\bm{\Omega}_-(\bm{k})$: red for positive and blue for negative. 
Thus, the red and blue spots in Fig.~\ref{fig2}(d) correspond to the 
WNs in Fig.~\ref{fig2}(c).

The monopole charge carried by the WN at $\bm{k}_\eta^\pm$ is 
$(1/2\pi)\iint_s \bm{\Omega}_-(\bm{k})\cdot d\bm{s}(\bm{k})={\rm sgn}
(\prod_\beta v_{\eta,\beta}^\pm)=\chi_\eta^\pm$, the chiralities of WNs. 
The integral is on a closed surface $s$ enclosing the WN in momentum space. 
The monopole charge is also identical to the Chern number of the lower magnon 
band on this surface. 
Thus topologically protected surface states exist between WNs. 
To see the surface states, we consider a slab whose two end 
surfaces are perpendicular to the [100] direction. The (100) surface BZ is 
represented by the yellow plane in Fig.~\ref{fig1}(b), where 
the projection of the high symmetry points of the first bulk 
BZ onto the first surface BZ are denoted by the barred symbols. 
The density plot of the spectral function on the top surface along the 
$\bar{\rm Z}\bar{\Gamma}\bar{\rm Z}$ line (a projection of both H'K'H' 
and HKH) are shown in Fig.~\ref{fig2}(e) and \ref{fig2}(f) for the model 
parameters used in Fig.~\ref{fig2}(b) and \ref{fig2}(c), respectively. 
The surface states with high density (red color) on the top surface 
between WNs can be clearly seen. 
Near the energy of WNs, these surface states form magnon arcs (an analogue of 
the Fermi arcs) on the sample surfaces, see Supplemental Material Ref. \cite{sm}.  

\begin{figure}[b]
  \begin{center}
  \includegraphics[width=8.5 cm]{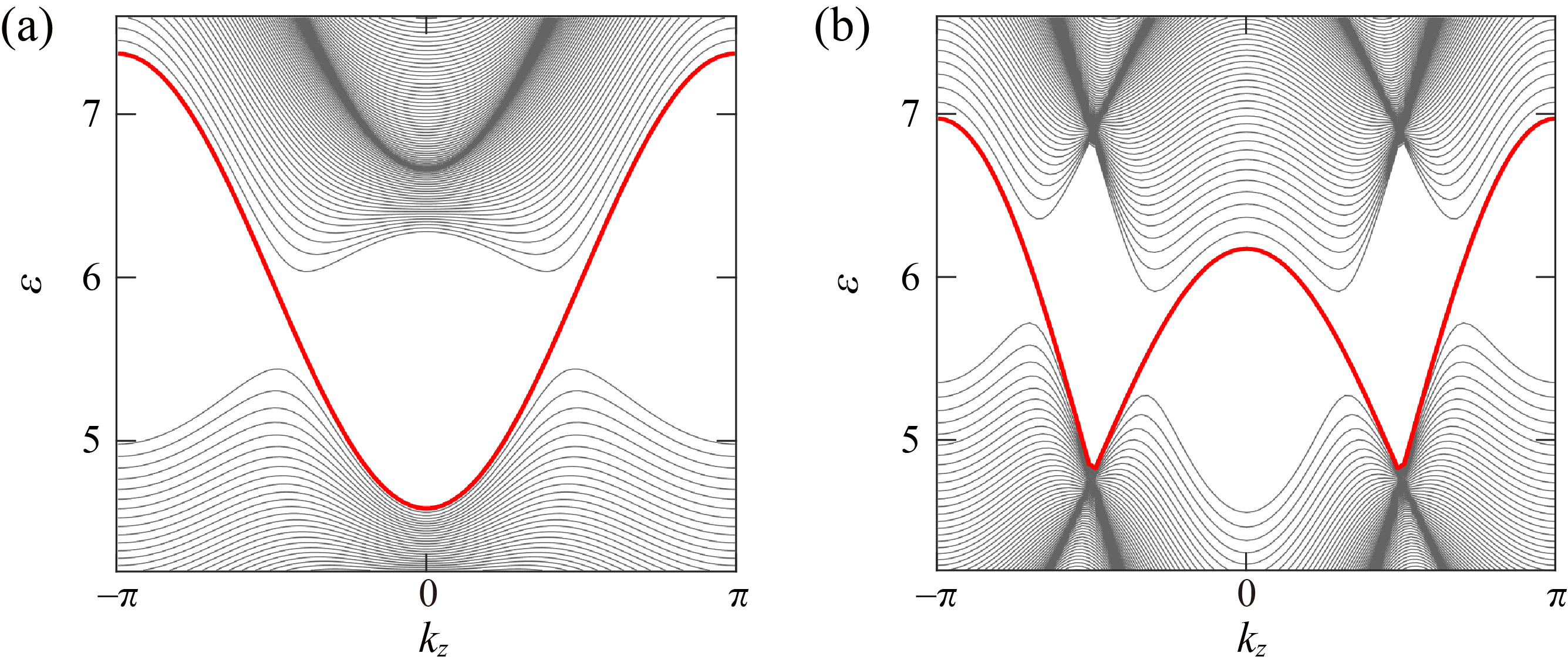}
  \end{center}
  \vspace{-0.2in}
\caption{(color online). \footnotesize{(a) and (b) Magnonic Landau levels 
under an electric field gradient of $\mathcal{E}=1/150\sqrt{3}$ (in units 
of $\hbar c^2/g\mu_B$) for the model parameters identical to those in 
Fig.~\ref{fig2}(b) and \ref{fig2}(c), respectively. The zeroth magnonic 
Landau levels are the red curves and the energy $\varepsilon$ is in 
units of $JS$.}}
  \label{fig3}
  \vspace{-0.2in}
\end{figure}

According to the Aharonov-Casher effect \cite{AC}, a magnon with a 
magnetic moment $\bm{\mu}=g\mu_B \hat{\bm{z}}$ interacts with an 
electric field $\bm{E}$ and acquires the Aharonov-Casher phase,  
$\phi_{ij}=(1/\hbar c^2)\int_j^i(\bm{E}\times\bm{\mu})\cdot d\bm{r}$. 
This effect is reminiscent of the magnetic field effect on electrons that 
induces the Aharonov-Bohm phase and leads to the Landau levels.
Indeed, the effect has already been used to generate the MLLs and magnonic quantum Hall effect \cite{MQH}. 
Here we consider magnons under an inhomogeneous electric field 
$\bm{E}=(\mathcal{E}x,0,0)$. Compare the Aharonov-Casher phase 
$\phi_{ij}=(g\mu_B/\hbar c^2)\int_i^j \mathcal{E}x dy$ for magnons 
with the usual Aharonov-Bohm phase for electrons, the lattice momentum 
in the effective Weyl Hamiltonian Eq.~(\ref{H3}) should be replaced by 
$-i\hbar\nabla+g\mu_B\mathcal{E}x \hat{\bm{y}}/c^2$ \cite{MQH}. 
The effective Weyl Hamiltonian in the electric field can be solved 
exactly and the magnon motion in the $xy$ plane is quantized into the 
MLLs with the eigenvalues 
\begin{equation}
\begin{split}
&\varepsilon_{\eta,n\geq 1}^\pm(q_z)=\varepsilon_{\eta}+
\hbar u_{\eta,z}^\pm q_z \pm \hbar \sqrt{\lambda_\eta n+
v_{\eta,z}^{\pm 2} q_z^2},\\
&\varepsilon_{\eta,0}^\pm(q_z)=\varepsilon_{\eta}+\hbar 
u_{\eta,z}^\pm q_z-(-1)^\eta 
{\rm sgn}(\mathcal{E})\hbar{v_{\eta,z}^{\pm} q_z},
\end{split}
\end{equation}
where $\lambda_\eta= 2|v_{\eta,x}^{\pm}v_{\eta,y}^{\pm}|/l_\mathcal{E}^2$ 
and electric length $l_\mathcal{E}=\sqrt{{\hbar c^2}/{g\mu_B |\mathcal{E}|}}$ 
is an analogue of magnetic length for electrons \cite{MQH}. 
The MLL degeneracy is $D=L_xL_y/2\pi l_\mathcal{E}^2\propto |\mathcal{E}|$, 
where $L_x$ and $L_y$ are the sample lengths in $x$ and $y$ directions. 
The zeroth MLL $\varepsilon_{\eta,0}^\pm(q_z)$ is chiral and linearly 
dispersed with opposite group velocities $v_{\eta,g}^\pm=u_{\eta,z}^\pm 
-(-1)^\eta{\rm sgn}(\mathcal{E}) v_{\eta,z}^\pm$ around two paired WNs at 
$\bm{k}_\eta^\pm$, where the density of states is  $\rho_\eta=
(2\pi^2l_\mathcal{E}^2\hbar|v_{\eta,g}^\pm|)^{-1}$ (with the Landau 
degeneracy included). We also include the Aharonov-Casher phase into the 
tight-binding Hamiltonian (\ref{H2}) through the Peierls substitution \cite{PS} 
and calculate its spectrum for an 
infinite long bar along $z$ direction with periodic boundary condition in $x$ 
and $y$ directions. For the electric field gradient $\mathcal{E}=1/150\sqrt{3}$ 
in units of $\hbar c^2/g\mu_B$, the MLLs for the same model parameters used in 
Fig.~\ref{fig2}(b) and \ref{fig2}(c) are shown in Fig.~\ref{fig3}(a) and 
\ref{fig3}(b), where the zeroth MLLs are the red curves. 
The MLLs for type-II WMs are shown in the Supplemental Material Ref.~\cite{sm}.


To realize the MCA, one needs to drive magnons 
to flow in the direction perpendicular to the magnon quantization plane. 
Due to the interaction energy $-\bm{B}\cdot \bm{\mu}$ between a magnetic 
field $\bm{B}$ and a magnetic moment $\bm{\mu}$, an inhomogeneous magnetic 
field of $\bm{B}\parallel \bm{\mu}$ can exert a force of $g\mu_B\partial_z B_z$ 
on a magnon so that magnon momentum shall follow the dynamical equation 
$\hbar dk_z/dt=g\mu_B\partial_z B_z$. The change of the magnon momentum 
drives magnons to flow from one WN to the other due to the unidirectional 
nature of the zeroth MLL. The transport of chiralities through the 
zeroth MLL leads to the non-conservation of chirality \cite{NN,CA}, an 
important feature of MCA. 
 
\begin{figure}
  \begin{center}
  \includegraphics[width=7.5 cm]{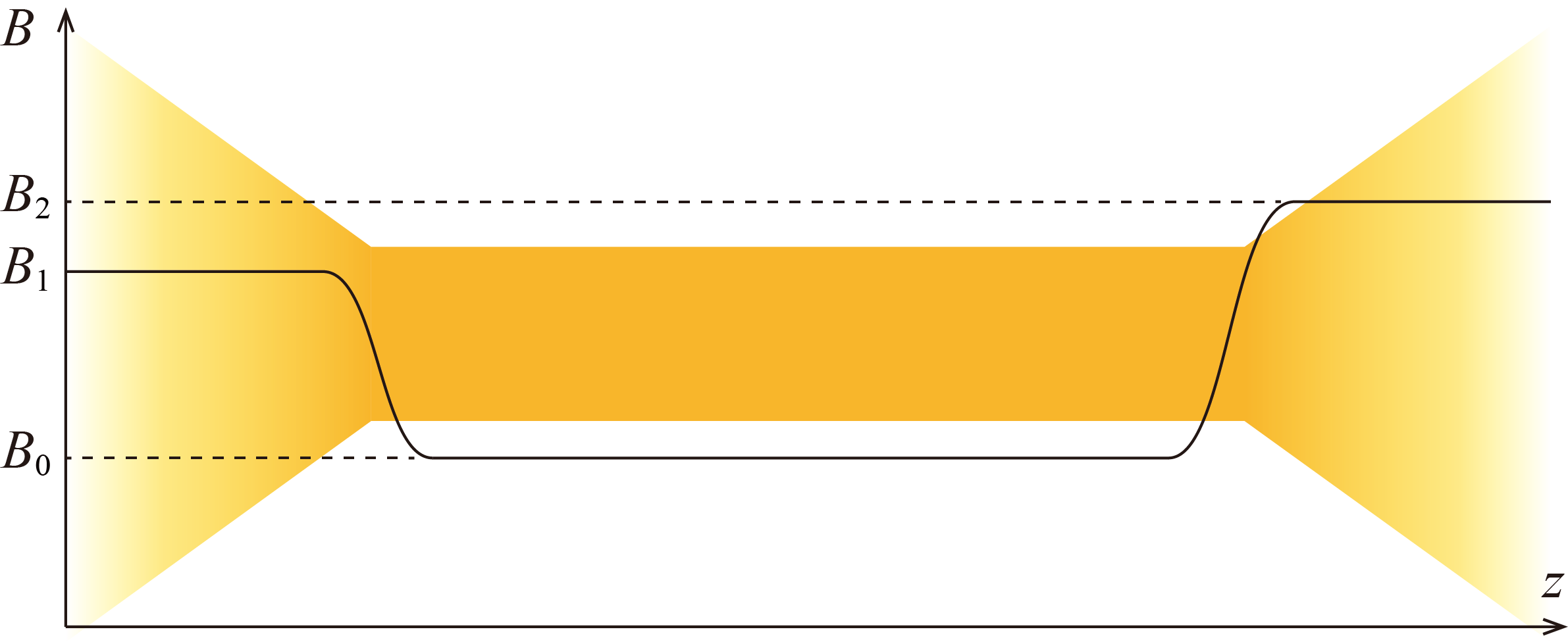}
  \end{center}
  \vspace{-0.2in}
\caption{(color online). \footnotesize{Schematic diagram of a two-terminal 
setup under an inhomogeneous magnetic field along $z$ direction, in which 
the middle quasi-one-dimensional magnon conductor is connected to two 
magnon reservoirs. }}
  \label{fig4}
  \vspace{-0.2in}
\end{figure}

To detect the MCA, we can consider a two-terminal setup sketched in Fig.~\ref{fig4} 
under an inhomogeneous magnetic field along the $z$ direction.  
Here a quasi-one-dimensional magnon conductor described by Eq.~(\ref{H1}) 
and in an inhomogeneous electric field described above is connected to 
two magnon reservoirs. Higher magnetic fields $B_1$ and $B_2$ are applied 
on the reservoirs to shift the magnon band bottom to 
$\varepsilon_\eta-g\mu_B \Delta B/2$ and $\varepsilon_\eta+g\mu_B \Delta B/2$ 
(where $\Delta B=B_2-B_1$) so that the system is at nonequilibrium. 
The imbalance of magnon concentrations between the two reservoirs within 
the energy window $[\varepsilon_\eta-g\mu_B\Delta B/2,\varepsilon_\eta+g\mu_B 
\Delta B/2]$ drives magnons to flow from the left to the right through the 
magnon conductor \cite{MQH}. In the ballistic regime where the sample length 
is smaller than the magnon mean free path, and for a type-I WM with only one 
pair of WNs at $\bm{k}^\pm_\eta$ ($\eta=1$ or 2), the spin and heat currents through 
the zeroth MLL can be calculated from the Landauer-B{\" u}ttiker approach 
\cite{MQH} as 
\begin{equation}
\begin{split}
&I_{s,\eta} 
=\int_{\varepsilon_\eta-g\mu_B\Delta B/2}^{\varepsilon_\eta+g\mu_B\Delta B/2} 
\frac{\hbar L_xL_yn_B(\varepsilon)d\varepsilon}{4\pi^2l_\mathcal{E}^2 \hbar}= 
G_{s,\eta}\Delta B,\\
&I_{h,\eta} =\int_{\varepsilon_\eta-g\mu_B\Delta B/2}^{\varepsilon_\eta+g\mu_B\Delta B/2} 
\frac{\varepsilon L_xL_y n_B(\varepsilon)d\varepsilon}{4\pi^2l_\mathcal{E}^2\hbar}= 
G_{h,\eta}\Delta B, \\
&G_{s,\eta}=\frac{L_xL_yg\mu_B n_B(\varepsilon_\eta)}{4\pi^2 l_\mathcal{E}^2},\;
G_{h,\eta}=\frac{L_xL_yg\mu_B\varepsilon_\eta n_B(\varepsilon_\eta)}
{4\pi^2 l_\mathcal{E}^2\hbar},
\end{split}
\end{equation}
where $n_B(\varepsilon)=(e^{\varepsilon/k_BT}-1)^{-1}$ is the Bose-Einstein 
distribution, $G_{s,\eta}$ and $G_{h,\eta}$ are respective the spin and heat 
conductance (from the pair of WNs labeled by $\eta$) in linear response. 
The neglect of the contributions from higher MLLs ($n\geq 1$) can be 
justified when $g\mu_B\Delta B\ll 2\hbar\sqrt{\lambda_\eta}$ such that 
the energy window is within the energy gap between the first MLLs 
$\varepsilon_{\eta,1}^\pm(q_z)$. Apparently, the spin (heat) conductance 
is linear in the electric field gradient 
$G_{s(h),\eta}\propto l_{\mathcal{E}}^{-2}\propto |\mathcal{E}|$ due to the 
MLL degeneracy. 
Therefore, the MCA results in positive and linear electric spin (heat) 
conductance, or negative electric spin (heat) resistance as 
$R_{s(h),\eta}\propto |\mathcal{E}|^{-1}$.
These results are experimentally detectable \cite{BT} and can be used as 
the signatures of MCA.  

Before concluding this paper, we would like to make the following remarks.
1) The above results should be valid for any number of pairs 
of nearly degenerate WNs as long as the WM is type-I because each pair of
WNs gives one zeroth MLL and different zeroth MLLs are additive. 
Thus, the total spin (heat) conductance is simply 
$G_{s(h)}\simeq\sum_{\eta} G_{s(h),\eta}$, due to different pairs of WNs 
labeled by $\eta$. 
The transport of type-II WMs can be complicated because their higher MLL 
channels also conduct magnons so that the zeroth MLL cannot be isolated from 
higher MLLs, see Supplemental Material Ref.\cite{sm}. The linear electric spin and 
heat conductance from MCA means that the electric field gradient can be used 
to control magnon transport, and this should open a new avenue for magnonics. 
For diffusive transport (when sample length is much larger than the mean 
free path), the electric field dependence of spin (heat) conductance 
should be sensitive to the detailed scattering processes. 
In fact, it was recently shown that the linear magnetoconductance 
can exist in disordered Weyl semimetals \cite{LM}. 
How does it works for WMs is an open question for future investigation. 
2) Besides the transport measurement, one can also study the WMs by 
examining WNs and magnon arcs detectable by inelastic neutron scattering that 
was successfully used to probe the magnon bands of a topological magnon
insulator \cite{TM12}.
3) There is clear difference between the MCA and its electronic counterpart, 
electronic chiral anomaly. Instead of the electric and magnetic fields 
parallel to each other, the inhomogeneous electric and magnetic fields 
in MCA must be perpendicular to each other. 

In conclusion, the stacked honeycomb ferromagnets can support both type-I 
and type-II WMs. MLLs can be realized by the interaction between electric 
field and magnon magnetic moment through the Aharonov-Casher effect. 
MCA results in linear dependence of spin and heat conductance on the electric 
field gradient when mutually perpendicular inhomogeneous electric and 
magnetic fields are applied.  Our results provide a new way to probe WMs 
and open a door to new electrically controlled magnonic devices. 

\begin{acknowledgements}
This work is supported by the NSF of China Grant (No. 11374249)
and Hong Kong RGC Grants (No. 163011151 and No. 16301816).
\end{acknowledgements}

\pagebreak
\widetext
\begin{center}
\textbf{\large Supplemental Material for Chiral Anomaly of Weyl Magnons in 
Stacked Honeycomb Ferromagnets}
\end{center}
\setcounter{equation}{0}
\setcounter{figure}{0}
\setcounter{table}{0}
\setcounter{page}{1}
\makeatletter
\renewcommand{\theequation}{S\arabic{equation}}
\renewcommand{\thefigure}{S\arabic{figure}}
\renewcommand{\bibnumfmt}[1]{[S#1]}
\renewcommand{\citenumfont}[1]{S#1}

{\it Chern number.}---In order to elaborate the existence of 
topologically protected surface states in both Weyl magnon (WM) and 
topological magnon insulator phases, we calculate the Chern number 
in the $k_xk_y$ plane from the Hamiltonian $\mathcal{H}(\bm{k})$ used 
in the manuscript for fixed $k_z$'s. 
The Chern number as a function of $k_z$ is
\begin{equation}
C(k_z)=\sum_{\eta=1,2}\frac{(-1)^\eta}{2}
{\rm sgn}\left[(-1)^\eta3\sqrt{3}D+K_-+J_-(1-\cos k_z)\right].
\label{C}
\end{equation} 
The Chern number changes from 0 to $\pm 1$ and vice versa when the 
constant $k_z$ plane pass through a Weyl node (WN) in momentum space. 
Therefore, the topologically protected surface states exist between 
WNs and the WNs are the starting or ending points of magnon arcs. 
$C(k_z)\neq 0$ for all $k_z$ in the topological magnon insulator phase,  
and topologically protected surface states exist in the magnon band gap.
 
\begin{figure}[b]
  \begin{center}
  \includegraphics[width=8.5 cm]{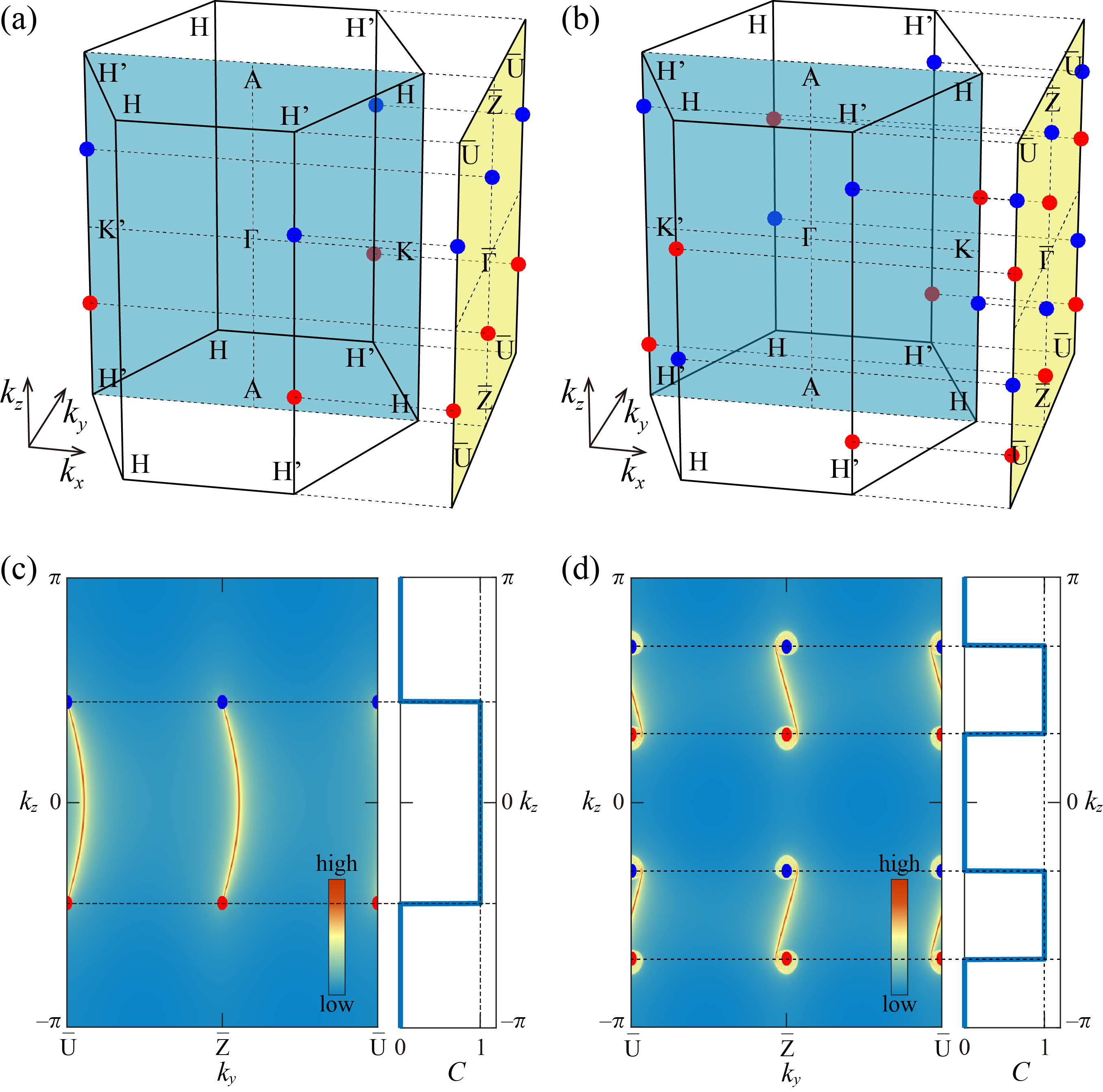}
  \end{center}
  \vspace{-0.2in}
\caption{\footnotesize{(a) and (b) The first bulk Brillouin zone (BZ) 
and the first (100) surface BZ of stacked honeycomb lattices. 
The red and blue dots denote the Weyl nodes (WNs) of chirality $\pm 1$ 
in Fig.~2(b) and 2(c) of the manuscript. (c) and (d) The density plot 
of the surface spectral functions on the first (100) surface BZ for 
model parameters used in Fig.~2(b) and 2(c) of the manuscript are shown 
in the left panels. The corresponding Chen numbers $C(k_z)$ are shown 
in the right panels. The red curves between WNs are magnon arcs. 
The insets are colorbars}}
  \label{s1}
  \vspace{-0.2in}
\end{figure}

{\it Magnon arc.---}
For model parameters used in Fig.~2(b) and 2(c) of the manuscript, WNs 
of chirality $\pm1$ in the first bulk Brillouin zone (BZ) and their 
projection in the first (100) surface BZ are respectively denoted by the 
red and blue dots in Fig.~\ref{s1}(a) and \ref{s1}(b), for the case of 
Fig.~2(b) and 2(c) in the manuscript. In order to see the magnon arcs 
formed from the topologically protected surface states near the WN energy 
on sample surfaces, we consider a slab of stacked honeycomb ferromagnets 
whose surfaces are perpendicular to the [100] direction, the same as that 
studied in the manuscript. For the model parameters used in Fig.~2(b) [or
2(e)] in the manuscript, there is one pair of WNs at $\bm{k}_1^\pm$ on 
H'K'H' in the first bulk Brillouin zone (BZ) as shown in Fig.~\ref{s1}(a).
For magnon energy $\varepsilon=\varepsilon_1$ through the WNs, the calculated 
(100) surface spectral function is shown in the left panel of Fig.~\ref{s1}(c). 
Apparently, the surface states with high density (red color) on the top surface 
form magnon arcs that connect the two WNs $\bm{k}_1^\pm$ of opposite chirality. 
The corresponding Chern number $C(k_z)$ is computed and shown in 
the right panel of Fig.~\ref{s1}(c). Indeed $C(k_{1,z}^-<k_z<k_{1,z}^+)=1$ 
is consistent with the appearance of magnon arcs as shown in Fig.~\ref{s1}(c).
For the model parameters used in Fig.2~(c) [or 2(f)] in the manuscript, there 
are two pairs of WNs at $\bm{k}_1^\pm$ on H'K'H' and at $\bm{k}_2^\pm$ on 
HKH in the first bulk BZ as shown in Fig.~\ref{s1}(b). Because the two 
pairs of WNs have difference energies $\varepsilon_1$ and $\varepsilon_2$, 
we set the magnon energy $\varepsilon=(\varepsilon_1+\varepsilon_2)/2$ in 
the middle of the two pairs of WNs. The calculated (100) surface spectral 
function is shown in the left panel of Fig.~\ref{s1}(d). In the current case, 
the magnon arcs exist between WNs $\bm{k}_1^+$ and  $\bm{k}_2^+$ of opposite 
chirality, and between WNs $\bm{k}_1^-$ and $\bm{k}_2^-$ of opposite chirality. 
The corresponding Chern number $C(k_z)$ is shown in the right panel of 
Fig.~\ref{s1}(d) where $C(k_{1,z}^-<k_z<k_{2,z}^-)=C(k_{2,z}^+<k_z<k_{1,z}^+)=1$ 
as expected.

\begin{figure}
  \begin{center}
  \includegraphics[width=12 cm]{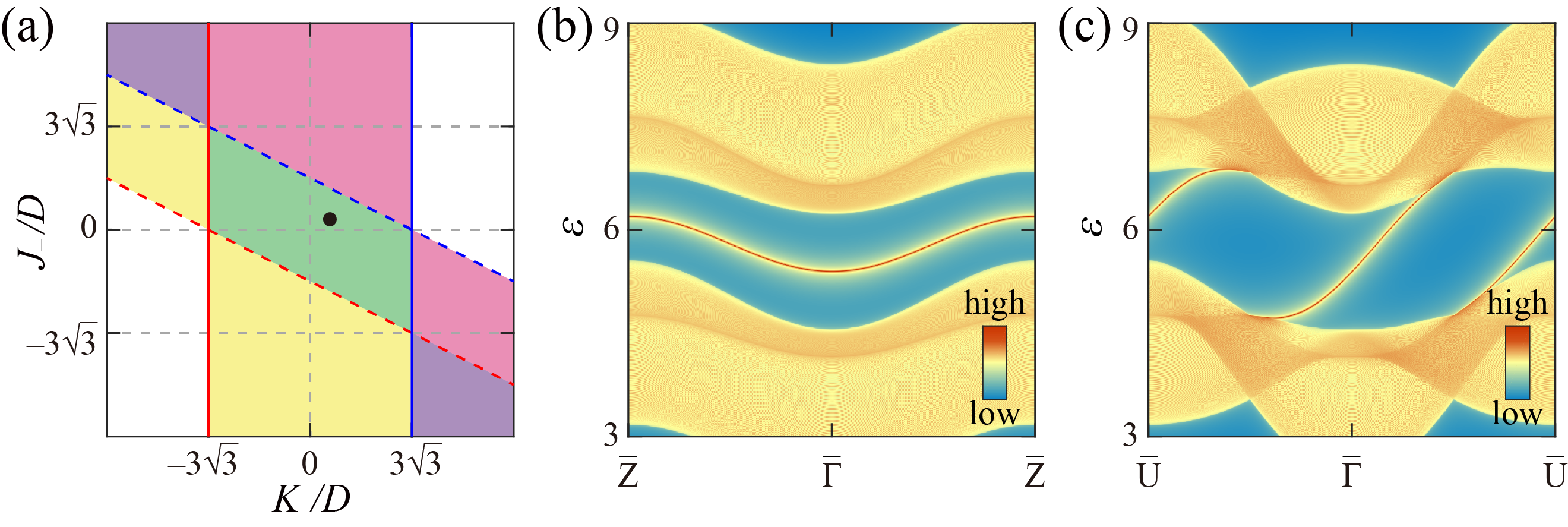}
  \end{center}
  \vspace{-0.2in}
\caption{(color online). \footnotesize{(a) Phase diagram of stacked honeycomb 
ferromagnets in the $K_-/D$-$J_-/D$ plane. (b) and (c) The density plot of the 
(100) surface spectral function for the topological magnon insulator phase 
[the black dot in (a)] along ${\rm\bar{Z}}{\rm\bar{\Gamma}}{\rm\bar{Z}}$ (b) 
and along ${\rm\bar{U}}{\rm\bar{\Gamma}}{\rm\bar{U}}$ (c). Energy $\varepsilon$ 
is in units of $JS$, and the insets are colorbars.}}
  \label{s2}
  \vspace{-0.2in}
\end{figure}

\begin{figure}[b]
  \begin{center}
  \includegraphics[width=8.5 cm]{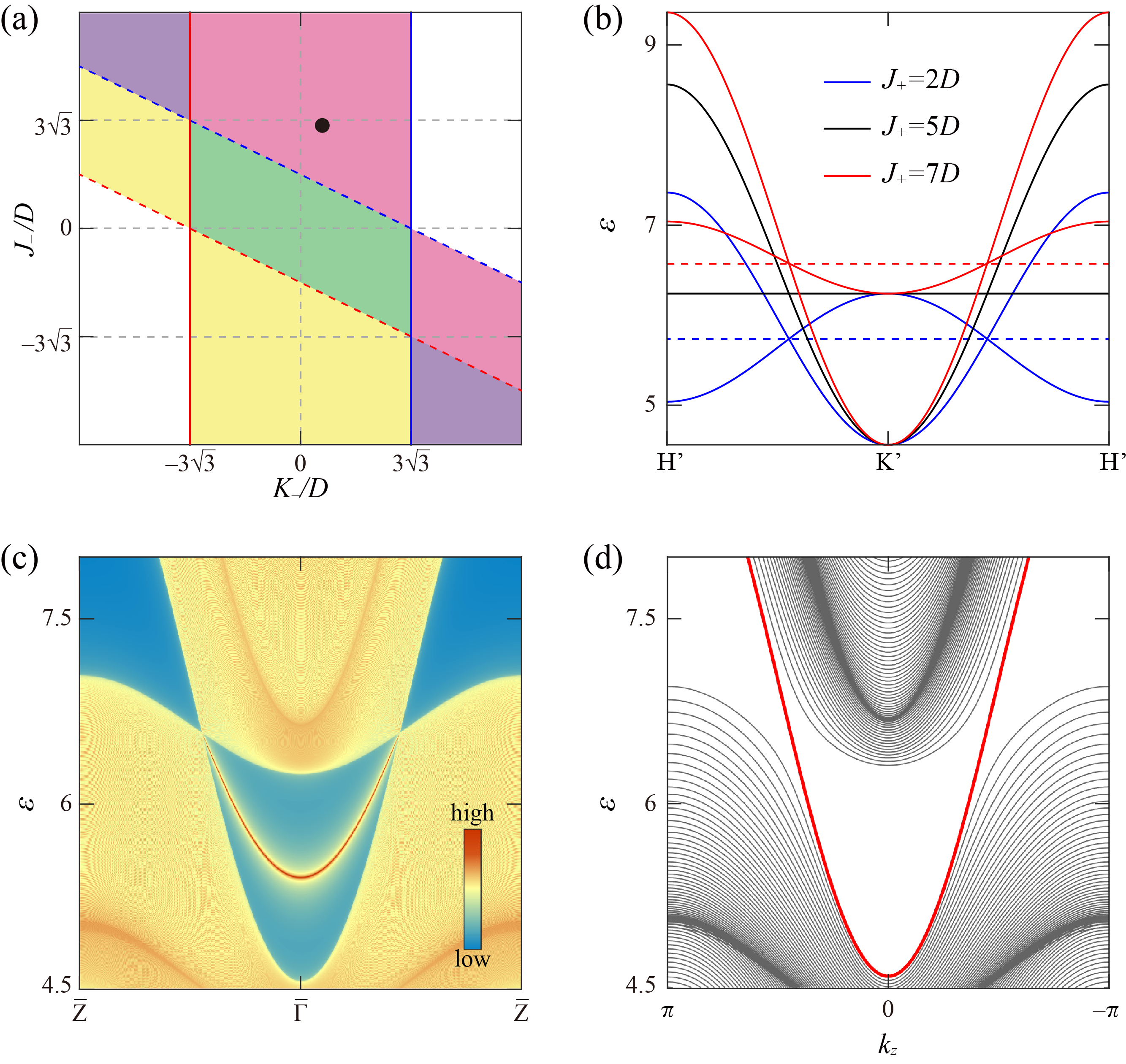}
  \end{center}
  \vspace{-0.2in}
\caption{(color online). \footnotesize{
(a) Phase diagram of stacked honeycomb ferromagnets in the $K_-/D$-$J_-/D$ plane. 
The various phases with different colors are explained in the manuscript. 
(b) Magnon bands for various $J_+$ and other parameters corresponding to the 
black dot in (a) are plotted along the H'K'H' line. 
(c) The density plot of the (100) surface spectral function for the type-II 
WM with $J_+=7D$ in (b) and along ${\rm\bar{Z}}{\rm\bar{\Gamma}}{\rm\bar{Z}}$. 
The inset is the colorbar. 
(d) The MLLs of the type-II WM under an inhomogeneous electric field 
which is specified in the text. The red curve is the zeroth MLL, and 
$\varepsilon$ is in units of $JS$.}}
  \label{s3}
  \vspace{-0.2in}
\end{figure}

\begin{figure}
  \begin{center}
  \includegraphics[width=8.5 cm]{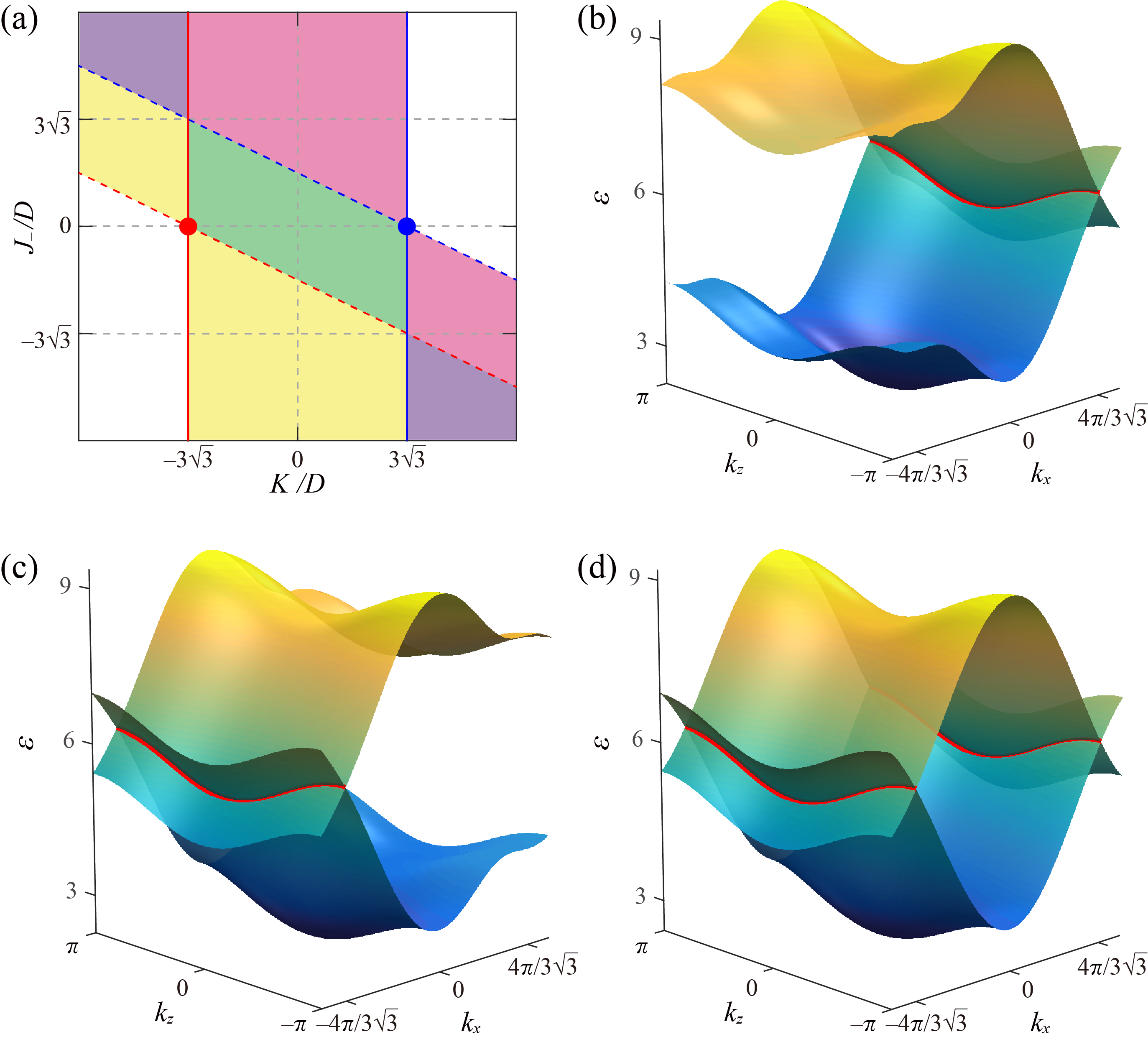}
  \end{center}
  \vspace{-0.2in}
\caption{ \footnotesize{(a) Phase diagram of stacked honeycomb ferromagnets 
in the $K_-/D$-$J_-/D$ plane. For the red and blue dots in (a), the magnon 
bands in the $k_xk_z$ plane are shown in (b) and (c) in which band 
crossing occurs on a nodal line marked by the red curve. 
(d) The magnon bands for $D=0$ and $K_-=0$ with other parameters specified 
in the text. The band crossing occurs on two nodal lines. 
$\varepsilon$ is in units of $JS$.}}
  \label{s4}
  \vspace{-0.2in}
\end{figure}

{\it Topological magnon insulator.---}According to Eq.~(\ref{C}), $C(k_z)=1$ in the topological magnon insulator phase 
[the Green region in Fig.~\ref{s2}(a)] for arbitrary $k_z$, while $C(k_z)=0$ 
in the trivial phase [the white regions in Fig.~\ref{s2}(a)] for arbitrary $k_z$.
To see the topologically protected surface states of the topological magnon 
insulator, we compute the (100) surface spectral function for the model 
parameters of $D=0.2J$, $K_+=12D$, $J_+=2D$, and $(K_-,J_-)=(D,0.5D)$ 
marked by the black dot in Fig.~\ref{s2}(a) inside the topological magnon 
insulator phase. The density plot of the (100) surface spectral function 
along ${\rm\bar{Z}}{\rm\bar{\Gamma}}{\rm\bar{Z}}$ and along 
${\rm\bar{U}}{\rm\bar{\Gamma}}{\rm\bar{U}}$ of the first (100) surface BZ 
are respectively shown in Fig.~\ref{s2}(b) and \ref{s2}(c). 
As expected, the topologically protected surface 
states exist in the magnon band gap. 

{\it Type-II WM and magnonic Landau level.---}In the manuscript we show that 
the WMs are of type-II when $|J_+|>|J_-|$ according to the classification 
of Weyl semimetals \cite{type2s}. This can be easily seen from the effective 
Weyl Hamiltonian Eq.~(3) of the manuscript in which the third term gives a 
Weyl cone while the second term tilts the Weyl cone along $k_z$ direction
in momentum space. According to the criteria of type-II Weyl semimetal in 
Ref.~\cite{type2s}, we obtain the condition for WMs becoming type-II in our 
model as $|u_{\eta,z}^\pm|>|v_{\eta,z}^\pm|\Rightarrow|J_+|>|J_-|$. 

To confirm this result, we compute the 
magnon bands for the model parameters $D=0.2J$, $K_+=12D$, $(K_-,J_-)=(D,5D)$ 
[indicated by the black dot in Fig.~\ref{s3}(a)], and $J_+=\{2D,5D,7D\}$.  
In the current case, the WNs are located on the H'K'H' line in the first bulk 
BZ as shown in Fig.~\ref{s1}(a). Therefore, we plot magnon energy along 
H'K'H' and the results are shown in Fig.~\ref{s3}(b). 
Indeed, the system changes from the type-I WM to the type-II WM at $J_+=5D=J_-$ 
where one magnon band is flat along the H'K'H' as shown in Fig.~\ref{s3}(b). 
For $J_+=7D>J_-$, it's clearly shown that the Weyl cones are tilted to be type-II.

Because the tilt of Weyl cone doesn't change its band topology, the topologically 
protected surface states should also exist between WNs in type-II WMs. 
In Fig.~\ref{s3}(c), we show the density plot of the (100) surface spectral 
function of the type-II WM with $J_+=7D$ in Fig.~\ref{s3}(b) 
along ${\rm\bar{Z}}{\rm\bar{\Gamma}}{\rm\bar{Z}}$ of the first (100) surface BZ. 
Indeed the the tilted Weyl cone and surface states (red curve) are clearly identified. 

According to the Aharonov-Casher effect \cite{ACs}, magnons can interact with 
electric fields that results in magnonic Landau levels (MLLs) as described  
in the manuscript. Here we consider the type-II WM coupled to an inhomogeneous 
electric field $\bm{E}=(\mathcal{E}{x},0,0)$ with $\mathcal{E}=1/100\sqrt{3}$ 
in units of $\hbar c^2/g\mu_B$. The MLLs of the type-II WM are shown in 
Fig.~\ref{s3}(d) and the zeroth MLL is the red curve. 
The zeroth MLL is still chiral but cannot be separated from higher MLLs, 
because there is no band gap between the first MLLs.  

{\it Nodal-line magnon.---}At $(K_-/D, J_-/D)=(-3\sqrt{3},0)$ [the red 
dot in Fig.~\ref{s4}(a)] and $(3\sqrt{3},0)$ (the blue dot), WMs becomes  
nodal-line magnons whose two magnon bands cross on HKH line for the 
former and on H'K'H' line for the later as mentioned in the manuscript.
In order to visualize these nodal lines, we calculate the magnon bands for the 
red and blue dots in Fig.~\ref{s4}(a) with the other model parameters $D=0.2J$, 
$K_+=12D$, and $J_+=2D$. The energy surfaces of two bands as a function of 
$k_x$ and $k_z$ for fixed $k_y=0$ [the blue plane in Fig.~\ref{s1}(a)] are 
plotted in Fig.~\ref{s4}(b) (for the red dot) and \ref{s4}(c) (for the blue dot). 
The nodal lines are marked by red curves. When $K_-=D=0$ (with the other 
parameters unchanged), the red and blue dots merge together and 
the two nodal lines coexist as shown in Fig.~\ref{s4}(d).


\begin{thebibliography}{99}
\bibitem{TM1}S. Fujimoto, Phys. Rev. Lett. \textbf{103}, 047203 (2009).
\bibitem{TM2}H. Katsura, N. Nagaosa, and P. A. Lee, Phys. Rev. Lett. \textbf{104}, 066403 (2010).
\bibitem{TM3}Y. Onose, T. Ideue, H. Katsura, Y. Shiomi, N. Nagaosa, and Y. Tokura,
Science \textbf{329}, 297 (2010).
\bibitem{TM4}R. Matsumoto and S. Murakami, Phys. Rev. Lett.
\textbf{106}, 197202 (2011);
R. Matsumoto, R. Shindou, and S. Murakami, Phys. Rev. B \textbf{89}, 054420 (2014).
\bibitem{TM5}L. Zhang, J. Ren, J.-S. Wang, and B. Li, Phys. Rev. B \textbf{87},
144101 (2013).
\bibitem{TM6}R. Shindou, J. I. Ohe, R. Matsumoto, S. Murakami, and
E. Saitoh, Phys. Rev. B \textbf{87}, 174402 (2013);
R. Shindou, R. Matsumoto,
S. Murakami, and J. I. Ohe, Phys. Rev. B \textbf{87}, 174427 (2013).
\bibitem{TM7}M. Mochizuki, X. Z. Yu, S. Seki, N. Kanazawa, W. Koshibae, 	
J. Zang, M. Mostovoy, Y. Tokura and N. Nagaosa, Nat. Mater. \textbf{13}, 241 (2014).
\bibitem{TM8}A. Mook, J. Henk, and I. Mertig, Phys. Rev. B \textbf{90}, 024412 (2014).
\bibitem{TM9}M. Mena, R. S. Perry, T. G. Perring, M. D. Le, S. Guerrero, M. Storni,
D. T. Adroja, Ch. R{\"u}egg, and D. F. McMorrow, Phys. Rev. Lett. \textbf{113}, 047202 (2014).
\bibitem{TM10}I. Lisenkov, V. Tyberkevych, A. Slavin, P. Bondarenko,
B. A. Ivanov, E. Bankowski, T. Meitzler, and S. Nikitov, Phys. Rev. B
\textbf{90}, 104417 (2014).
\bibitem{TM11}H. Lee, J. H. Han, and P. A. Lee, Phys. Rev. B \textbf{91}, 125413 (2015).
\bibitem{TM12}R. Chisnell, J. S. Helton, D. E. Freedman, D. K. Singh, R. I. Bewley,
D. G. Nocera, and Y. S. Lee, Phys. Rev. Lett. \textbf{115}, 147201 (2015).
\bibitem{TM13}C.-E. Bardyn, T. Karzig, G. Refael, and T. C. H. Liew, Phys. Rev. B \textbf{93}, 020502(R) (2016).
\bibitem{TM14}S. A, Owerre, J. Phys.: Condens. Matter \textbf{28}, 386001 (2016).
\bibitem{TM15}R. Cheng, S. Okamoto, and D. Xiao, Phys. Rev. Lett. {\bf 117}, 217202 (2016).
\bibitem{TM16}V. A. Zyuzin and A.A. Kovalev, Phys. Rev. Lett. {\bf 117}, 217203 (2016).
\bibitem{TM17}S. K. Kim, H. Ochoa, R. Zarzuela, and Y. Tserkovnyak, Phys. Rev. Lett. {\bf 117}, 227201 (2016).
\bibitem{DM}J. Fransson, A. M. Black-Schaffer, and A. V. Balatsky, Phys. Rev. B {\bf 94}, 075401 (2016).
\bibitem{WM1}F. Y. Li, Y. D. Li, Y. B. Kim, L. Balents, Y. Yu, and  G. Chen, Nature Commun. {\bf 7}, 12691 (2016).
\bibitem{WM2}A. Mook, J. Henk, and I. Mertig, Phys. Rev. Lett. {\bf 117}, 157204 (2016).
\bibitem{WM3}Y. Su, X. S. Wang, and X. R. Wang, arXiv:1609.01500 (2016).
\bibitem{NL}A. Mook, J. Henk, and I. Mertig, Phys. Rev. B. {\bf 95}, 014418
(2017).
\bibitem{TM19}X. S. Wang, Y. Su, and X. R. Wang, Phys. Rev. B. {\bf 95}, 014435 (2017).

\bibitem{M1}S. O. Demokritov, and A. N. Slavin, {\it Magnonics: From
Fundamentals to Applications}
(Topics in Applied Physics Vol. 125, Springer, Berlin, 2013).
\bibitem{M2}D. Grundler, Nat. Phys. \textbf{11}, 438 (2015).
\bibitem{M3}B. Lenk, H. Ulrichs, F. Garbs, and M. Münzenberg,
Phys. Rep. \textbf{507}, 107 (2011).
\bibitem{M4}A. V. Chumak, V. I. Vasyuchka, A. A. Serga, and B. Hillebrands,
Nat. Phys. \textbf{11}, 453 (2015).
\bibitem{TS1}Y. Kajiwara, K. Harii, S. Takahashi, J. Ohe, K. Uchida,
M. Mizuguchi, H. Umezawa, K. Kawai, K. Ando, K. Takanashi,
S. Maekawa, and E. Saitoh, Nature (London) {\bf 464}, 262 (2010).
\bibitem{TS2} L. J. Cornelissen, J. Liu, R. A. Duine, J. B. Youssef, and B. J.
van Wees, Nature Phys. {\bf 11}, 1022 (2015).
\bibitem {yanpeng}P. Yan, X. S. Wang, and X. R. Wang, Phys. Rev. Lett.
\textbf{107}, 177207 (2011).
\bibitem{xiansi}X. S. Wang, P. Yan, Y. H. Shen, G. E. W. Bauer, and
X. R. Wang, Phys. Rev. Lett. \textbf{109}, 167209 (2012).
\bibitem {hubin}B. Hu and X. R. Wang, Phys. Rev. Lett.
\textbf{111}, 027205 (2013).
\bibitem{AC}Y. Aharonov and A. Casher, Phys. Rev. Lett. {\bf 53}, 319 (1984).
\bibitem{ab}The various in-plane vectors in Fig.~\ref{fig1}(a) are set as 
$\bm{a}_1=(0,-1,0)$, $\bm{a}_2=(\sqrt{3}/2,1/2,0)$, $\bm{a}_3=(-\sqrt{3}/2,{1}/{2},0)$, 
$\bm{b}_1=(\sqrt{3}/{2},-{3}/{2},0)$, $\bm{b}_2=(\sqrt{3}/{2},{3}/{2},0)$, and 
$\bm{b}_3=(-\sqrt{3},0,0)$, where the distance 
between nearest neighbor intralayer lattice sites is set as unity.
\bibitem{DMI}I. Dzyaloshinskii, J. Phys. Chem. Solids {\bf 4}, 241 (1958); T.
Moriya, Phys. Rev. 120, {\bf 91} (1960).
\bibitem{HPT}T. Holstein and H. Primakoff, Phys. Rev. {\bf 58}, 1098
(1940).
\bibitem{sm} See Supplemental Material for magnon arcs, magnon bands, and surface spectral functions for the topological magnon insulator, type-II WM, and nodal-line magnon.
\bibitem{C}P. Hosur and X. L. Qi, C. R. Physique 14, 857870 (2013).
\bibitem{NN}H. B. Nielsen and M. Ninomiya, Phys. Lett. B {\bf 130},
389 (1983).
\bibitem{WAN}X. Wan, A. M. Turner, A. Vishwanath, and S. Y.
Savrasov, Phys. Rev. B {\bf 83}, 205101 (2011).
\bibitem{T2WM}A. A. Soluyanov, D. Gresch, Z. Wang, Q. Wu, M. Troyer, X. Dai, 
and B. A. Bernevig, Nature (London) {\bf 527}, 495 (2015).
\bibitem{MQH}F. Meier and D. Loss, Phys. Rev. Lett. {\bf 90}, 167204 (2003); 
K. Nakata, J. Klinovaja, and D. Loss, arXiv:1611.09752 (2016). 
\bibitem{PS}X. R. Wang, Phys. Rev. B  {\bf 53}, 12035 (1996).
\bibitem{CA}A. Burkov, Science 350, 378 (2015).  
\bibitem{BT}S. Y. Li, L. Taillefer, C. H. Wang, and X. H. Chen, Phys. Rev. Lett. {\bf 95}, 156603 (2005).
\bibitem{LM}H. Z. Lu and S. Q. Shen, Front. Phys. {\bf 12}, 127201 (2017).

\end{thebibliography}

\begin{thebibliography}{99}
\bibitem{type2s}
A. A. Soluyanov, D. Gresch, Z. Wang, Q. Wu, M. Troyer, X. Dai, 
and B. A. Bernevig, Nature (London) {\bf 527}, 495 (2015). 
\bibitem{ACs}
Y. Aharonov and A. Casher, Phys. Rev. Lett. {\bf 53}, 319 (1984). 
\end{thebibliography}
\end{document}